\documentclass[aps,prb,twocolumn,superscriptaddress]{revtex4-2}
\usepackage{bm}
\usepackage{graphicx}
\usepackage{graphics}
\usepackage{amsmath}
\usepackage{amsfonts}
\usepackage{amssymb}
\usepackage{makeidx}
\usepackage[colorlinks=true,citecolor=blue,linkcolor=blue,linktocpage=true,pagebackref=false]{hyperref}
\hypersetup{colorlinks=true,citecolor=blue,linkcolor=blue,filecolor=blue,urlcolor=blue}
\usepackage{color,soul} 
\usepackage{float}
\usepackage{subfigure}

\usepackage{siunitx}

\begin{document}

\newcommand{\be}   {\begin{equation}}
\newcommand{\ee}   {\end{equation}}
\newcommand{\ba}   {\begin{eqnarray}}
\newcommand{\ea}   {\end{eqnarray}}
\newcommand{\ve}   {\varepsilon}
\newcommand{\Dis}  {\mbox{\scriptsize dis}}

\newcommand{\state} {\mbox{\scriptsize state}}
\newcommand{\band} {\mbox{\scriptsize band}}

\title{High-energy Landau levels in graphene beyond nearest-neighbor hopping processes: \\ Corrections to the effective Dirac Hamiltonian}

\author{Kevin J. U. Vidarte}
\affiliation{Instituto de F\'{\i}sica, Universidade Federal
  do Rio de Janeiro, 21941-972 Rio de Janeiro - RJ, Brazil}
\author{Caio Lewenkopf}
\affiliation{Instituto de F\'{\i}sica, Universidade Federal
  Fluminense, 24210-346 Niter\'oi - RJ, Brazil}
\affiliation{Max Planck Institute for the Physics of Complex Systems, 01187 Dresden, Germany}

\date{\today}
\begin{abstract}
We study the Landau level spectrum of bulk graphene monolayers beyond the Dirac Hamiltonian with linear dispersion.
We consider an effective Wannier-like tight-binding model obtained from {\it ab initio} calculations, that includes long-range electronic hopping integral terms. 
We employ the Haydock-Heine-Kelly recursive method to numerically compute the Landau level spectrum of bulk graphene in the quantum Hall regime and demonstrate that this method is both accurate and computationally much faster than the standard numerical approaches used for this kind of study. 
The Landau level energies are also obtained analytically for an effective Hamiltonian that accounts for up to third nearest neighbor hopping processes.
We find an excellent agreement between both approaches. 
We also study the effect of disorder on the electronic spectrum. 
Our analysis helps to elucidate the discrepancy between theory and experiment for the high-energy Landau levels energies. 
\end{abstract}
\maketitle

\section{Introduction}
\label{sec:introduction}

The pioneering experimental reports \cite{Novoselov2005,Zhang2005} on the unique features of the quantum Hall (QH) effect in graphene systems played a key role to establish the remarkable massless Dirac electronic properties of this material \cite{CastroNeto2009, Goerbig2011, Yin2017}.
Subsequent investigations on QH effects in graphene continued to produce fascinating results, such as Klein tunneling \cite{Williams2007,Carmier2010}, fractional quantum Hall effect \cite{Du2009, Dean2011}, Hofstadter butterflies \cite{Dean2013, Hunt2013}, to name a few, attracting a lot of attention to the field (see, for instance, Ref.~\cite{Goerbig2022} for a recent review).  

In addition to  electronic transport properties, graphene under strong magnetic fields also shows unique spectral properties under strong magnetic fields. The graphene Landau levels have been  theoretically predicted \cite{McClure1956,Goerbig2011} to follow
\be
\label{eq:LL_naive}
\epsilon_N = {\rm sgn}(N) \hbar \omega_c \sqrt{\vert N\vert} 
\ee
where $N$ is the Landau level index, $\omega_c = v_F \sqrt{2 e B/\hbar}$ is the cyclotron frequency, $B$ is the magnetic field strength, and $v_F$ stand for the electron Fermi velocity 
for $B=0$.

The LLs spectrum have been experimentally measured by transmission \cite{Sadowski2006, Jiang2007} and scanning tunneling spectroscopy \cite{Li2009, Song2010} for both exfoliated and epithaxial graphene. These studies have verified the $\sqrt{BN}$ dispersion to a good approximation. A systematic transmission spectroscopy study \cite{Plochocka2008} focused on the high-energy LL showed deviations from $\epsilon_N$ predicted by Eq.~\eqref{eq:LL_naive}. The puzzle is that the disagreement with the experimental data persists even when one improves the theoretical description by accounting next-to-nearest neighbor hopping processes. 

In should be stressed that, while the massless Dirac Hamiltonian has been shown to be very effective in describing a variety of low energy electronic graphene properties, the theoretical modeling for higher energies is not unique. 
The standard approach uses density functional theory (DFT) calculations to obtain a tight-binding model based on Wannier orbitals, that contain an arbitrary range of hopping terms that fit the nonlinear features of the dispersion relation \cite{Charlier1991, Reich2002, Urban2011, Lherbier2012, Jung2013, Linhart2018}.  

Another possibility for the discrepancy between theory and experiment is disorder, which is ubiquitous in graphene \cite{Mucciolo2010}. 
Numerical investigations have studied the broadening the Landau subbands \cite{Zhu2009}, but there is no systematic study of the corresponding disorder-induced peak shifts. We examine this issue with emphasis on the large $|N|$ limit.

Our analysis uses the Haydock-Heine-Kelly (HHK) recursive method \cite{Haydock1972,Haydock1975,Haydock1980}, also called the Haydock method, an order ${\cal N}$ \cite{Fan2021} real-space computational approach developed to study local spectral functions. 
It transforms an arbitrary sparse Hamiltonian matrix in a tridiagonal form and evaluates the diagonal Green's function by a continued fraction expansion, avoiding the need of solving the full eigenvalue problem. 
The HHK method has been successfully used to compute the LDOS of different compounds \cite{Hsiao1988,Woodruff1987} and more recently in the study of carbon nanotubes \cite{Busser2013, Pincak2013}, and disordered graphene systems \cite{Shimomura2011,Smotlacha2012,Eremenko2016}.

In addition to its efficiency, another attractive feature of the HHK method is that, since it relies on the concept of nearsightedness \cite{Prodan2005}, it does not use periodic boundary conditions \cite{Haydock1980b}. 
Hence, it can be applied to study local spectral properties of disordered systems, quasicrystals \cite{Ahn2018, Yao2018}, and systems with very large primitive unit cells, such as twisted graphene layers and 2D systems under realistic magnetic fields $B$. 
Surprising the HHK method has only been employed once in a case where $B\neq 0$, namely, in the investigation of Hoftstadter butterfly energy gaps in square lattices \cite{Czycholl1988}. To the best of our knowledge, so far no one has realized that the method is also fast and very accurate (as we show) for the study of discrete QH spectra.

Regarding our results, we numerically show that the large-$|N|$ LL spectrum analytical solution of the continuum (long wavelength) effective graphene Hamiltonian is very accurate up to $\vert N \vert \leq 25$ and $B=25$~T.
We include DFT-fitted third nearest hopping matrix elements into the graphene tight-binding model Hamiltonian and show that agreement between theory and experiment is significantly improved.

This paper is organized as follows. In Sec.~\ref{sec:method}, we present the model Hamiltonian we use to describe the electronic properties in graphene, namely, a tight-binding Hamiltonian that accounts for up to third nearest neighbor hopping processes.
Next, we discuss disorder effects and review the SCBA predictions for the shifts in the LL subband peak energies and widths. 
Finally, we briefly present the main steps of the implementation of the HHK method, discuss its computational cost and benchmark its accuracy. 
In Sec.~\ref{sec:results} we present our results. We expand previous analytical results for $\epsilon_N$ and show that, despite the approximations involved (discussed in App.~\ref{sec:appendix}), the agreement with the numerical values is remarkable. 
We further show that the inclusion of third nearest neighbor matrix elements helps to improve the agreement between theory and experiment and that disorder plays a minor role. We summarize our conclusions in Sec.~\ref{sec:conclusion}.

\section{Theory and methods}
\label{sec:method}

\subsection{Model Hamiltonian}

The tight-binding Hamiltonian that describes the electronic structure of graphene monolayers reads \cite{Reich2002,CastroNeto2009}
\be
\label{eq:H_tb}
H = \sum_{i,j} \left(  t_{ij} c^\dagger_i c^{}_j + {\rm H. c} \right) ,
\ee
where $t_{ij}$ stands for the hopping matrix element between the Wannier electronic orbitals centered at the carbon sites $i$ and $j$.
Most studies consider only first nearest neighbor hopping processes, a simple model that is able to describe the low energy properties of bulk graphene \cite{CastroNeto2009,Mucciolo2010}.
Tight-binding parameterizations based on density functional theory (DFT) \cite{Charlier1991,Lherbier2012,Linhart2018,Urban2011} show the necessity to including hopping terms beyond first nearest neighbors for a more accurate modeling of the electronic dispersion, particularly when addressing higher energies. Here, we consider first $t^{(1)}$, second $t^{(2)}$, and third $t^{(3)}$ nearest neighbor hopping terms. Within this approximation it is convenient to write the graphene Hamiltonian in a sublattice matrix representation that in reciprocal space reads \cite{Bena2009, Goerbig2011}
\be
\label{eq:Hamiltonian matrix}
H_{\textbf{k}}\equiv \left( 
\begin{array}{cc}
t^{(2)}\vert \gamma_{\textbf{k}}\vert^{2} & t^{(1)}\gamma_{\textbf{k}}^{*}+t^{(3)}			
\gamma_{\textbf{k}}' \\ 
t^{(1)}\gamma_{\textbf{k}}+t^{(3)}\gamma_{\textbf{k}}'^{*} & t^{(2)} \vert \gamma_{\textbf{k}}
\vert^{2} 
\end{array} 
\right) ,
\ee
with
\be
\label{eq:sum of the NN(3N) phase factors}
\gamma_{\textbf{k}} \equiv 1 + e^{i \textbf{k} \cdot \textbf{a}_{2}}+
e^{i \textbf{k} \cdot (\textbf{a}_{2}-\textbf{a}_{1})}
\ee
and
\be
\label{eq:sum of the 4N phase factors}
\gamma '_{\textbf{k}} \equiv 1+ e^{i 2 \textbf{k} \cdot \textbf{a}_{2}}+
e^{i 2 \textbf{k} \cdot (\textbf{a}_{2}-\textbf{a}_{1})} ,
\ee
where $\textbf{a}_{1}=\sqrt{3} a_0 \hat{\bf e}_{x}$ and $\textbf{a}_{2}=\sqrt{3}a_0/2 \left( \hat{\bf e}_{x} + \sqrt{3}\hat{\bf e}_{y} \right)$ are the primitive vectors honeycomb lattice and $a_0=1.4 \,\si{\angstrom}$ is the carbon-carbon bond length \cite{CastroNeto2009}. 
The first and third nearest neighbor hopping terms, that connect different sublattices  \cite{CastroNeto2009,Goerbig2011}, correspond to the off-diagonal matrix elements, while the second-nearest hoppings are related to the diagonal ones. This apparent correspondence between even-odd nearest neighbors and diagonal off-diagonal matrix elements breaks down for fourth nearest neighbors and beyond \cite{Jung2013}. In Eq.~\eqref{eq:Hamiltonian matrix} we neglect a constant diagonal term that shifts the energy spectrum by $-3t^{(2)}$.

The energy dispersion reads
\be 
\label{eq:Energy dispersions}
\epsilon_{{\bf k}\lambda}=t^{(2)}\vert \gamma_{\textbf{k}}\vert^{2} + \lambda  \vert
t^{(1)}\gamma_{\textbf{k}}+t^{(3)} \gamma_{\textbf{k}}'^{*}\vert,
\ee
where $\lambda$ labels the valence ($\lambda=-1$) and conduction bands ($\lambda=+1$). 
Note that the second-neighbors hopping contributions do not depend of $\lambda$ and, thus, break the electron-hole symmetry.

The presence of an external magnetic field $\textbf{B}$ can be accounted by the Peierls substitution \cite{Peierls1933, Cresti2021}, that is, by the transformation
\be 
\label{eq:Peierls}
t_{ij} \longrightarrow t_{ij} \, \exp \left[ i \dfrac{e}{\hslash} \int_{\textbf{R}_{i}}^
{\textbf{R}_{j}} d\textbf{r}\cdot \textbf{A}(\textbf{r}) \right] ,
\ee
where $\textbf{R}_{n}$ is the lattice vector associated with the site $n$, $e$ is the electron charge. Here, we choose the vector potential $\textbf{A}(\textbf{r})= \left( 0,Bx,0\right)$ that gives a magnetic field $\textbf{B}=\nabla\times \textbf{A} = B\hat{\bf e}_{z}$ perpendicular to the graphene layer.

\subsection{Numerical method}
\label{sec:numerical_method}

We compute the Landau level spectra using the Haydock-Heine-Kelly (HHK) recursion technique \cite{Haydock1972, Haydock1975, Haydock1980}. 
The latter has been developed to calculate the local properties of electronic systems represented on a basis of localized (orthogonal) states ${\vert i\rangle}$, like the one used in the tight-binding Hamiltonian of Eq.~(\ref{eq:H_tb}). 
The HHK method provides a very efficient $O({\cal N})$ recursive procedure to transform a given Hamiltonian matrix into a tridiagonal one, that is much more amenable for numerical calculation. 
As mentioned introduction, the HHK method has been successfully used to compute the LDOS of continuous spectra of several systems \cite{Woodruff1987, Hsiao1988, Shimomura2011, Smotlacha2012, Busser2013, Pincak2013, Eremenko2016}. 
Here, we use it to compute the LDOS of a Hamiltonian with discrete eigenvalues.

Let us quickly review the main ingredients of the HHK method before discussing its use in computing the graphene Landau level spectrum.

The recursion method starts by targeting a given state $\vert 0 \rbrace = \vert j \rangle$. The method generates a hierarchy of states $\vert n \rbrace $ based on the three-term recursion relations \cite{Haydock1972, Haydock1975, Haydock1980}
\begin{align}
\label{eq:Hamiltonian_TD}
H\vert n\rbrace = a_{n}\vert n\rbrace +b_{n}\vert n-1\rbrace + b_{n+1}\vert n+1\rbrace,
\end{align}
with the recursive coefficients 
\begin{align}
\label{eq:recursive_coefficients_a}
a_{n}=\lbrace n \vert H\vert n\rbrace,
\end{align}
\begin{align}
\label{eq:recursive_coefficients_b}
b_{n+1}=\Vert (H-a_{n})\vert n\rbrace - b_{n}\vert n-1\rbrace \Vert,
\end{align}
and the orthogonal basis element
\begin{align}
\label{eq:orthogonal_basis}
\vert n+1\rbrace=\frac{1}{b_{n+1}} \left[  (H-a_{n})\vert n\rbrace - b_{n}\vert n-1\rbrace \right],
\end{align}
where $b_{0}=0$ and $\vert -1 \rbrace = 0$. 
Thus, by construction, the Hamiltonian matrix in the orthogonal basis $ \lbrace  \vert n \rbrace \rbrace $  is tridiagonal. 
In turn, the basis functions $\vert n\rbrace$ can be expressed in terms of Wannier-like states $\vert j \rangle $,
\begin{align}
\label{eq:the basis functions}
\vert n\rbrace = \sum_{i=1}^{P} A_{ni} \vert i\rangle .
\end{align}
Here, we assume that the electronic wave functions are a superposition of $P$ states centered at the atomic sites $i$ and the atomic orbitals are orthogonal to each other, in line with the tight binding model of Eq.~\eqref{eq:H_tb}. 

The diagonal Green's function for the seed state $\vert 0 \rbrace$ is given by continued fraction 
\cite{Haydock1972, Haydock1975, Haydock1980} 
\begin{align}
\label{eq:continuous_fraction}
\begin{split}
G_{00}(\epsilon) & =\lbrace 0 \vert \dfrac{1}{\epsilon-H} \vert 0 \rbrace \\
& =\dfrac{1}{\epsilon-a_{0}-{\textstyle \dfrac{b_{1}^{2}}{\epsilon-a_{1}-\dfrac{b_{2}^{2}}{\epsilon-a_{2}-\dfrac{b_{3}^{2}}{\ddots}}}}},
\end{split}
\end{align}
expressed in terms of the matrix elements of the tridiagonal Hamiltonian in the basis $\vert n \rbrace $. 
The LDOS at any site $j$ can be written as
\begin{align}
\label{eq:LDOS}
{\rm LDOS}(j,\epsilon)&= -\frac{1}{\pi} {\rm Im}\;G^{r}_{jj}(\epsilon) \\ \nonumber
&\equiv -\frac{1}{\pi} \lim_{\eta \rightarrow 0^{+}} \left[ {\rm Im}\;G_{jj}(\epsilon+i\eta) \right] .
\end{align} 
In practice, a finite $\eta$ serves as a convenient regularization parameter.
For continuous spectra, setting $\eta \approx 2D/M$, where $D$ is the bandwidth, guarantees a nice smooth approximation to LDOS$(j, \epsilon)$ \cite{Haydock1972}.

For pristine systems, due to translational symmetry, the LDOS$(j,\epsilon)$ at any $j$ is proportional to the (total) density of states $\rho(\epsilon)$ 
\footnote{Here, LDOS$(j,\epsilon)$ is normalized to unit and $\rho(\epsilon)$ to $2{\cal A}_{BZ}$ where ${\cal A}_{BZ}$ is the Brillouin zone area.}.
In our calculations we fix $j$ at the center of the honeycomb lattice of size $P$. 

In previous applications \cite{Czycholl1988,Hsiao1988,Woodruff1987} it has been observed that 
for a sufficiently large $M$, the recursive coefficients converge towards their asymptotic values, namely, $a_{M}\rightarrow a_{\infty}$ and $b_{M}\rightarrow b_{\infty}$. 
The asymptotic value of $a_{\infty}$ is associated with the center of the energy band $\epsilon_{0}$, namely, $\epsilon_{0}= a_{\infty}/2$ \cite{Haydock1972,Haydock1975}. 
In graphene systems, $a_{\infty}/2$ corresponds to the Dirac point energy. 
For the nearest-neighbor tight-binding Hamiltonian, the center band energy is $\epsilon_{0}=0$ and all $a_{n}$ coefficients are zero. 
In this case, by inspecting Eq.~\eqref{eq:continuous_fraction}, one immediately finds that $G_{00}(\epsilon)=-G_{00}(-\epsilon)$.
This implies that the LDOS$(j, \epsilon)$ has electron-hole symmetry for any $j$ and $\rho(\epsilon_0) = 0$, a condition that defines the so-called Dirac points \cite{CastroNeto2009}. 
When $t^{(2)}\neq 0$, the coefficient $a_{n}$ are no longer zero, the Dirac points are energy shifted, and the electron-hole symmetry broken. 
These simple properties are in line with well established literature results \cite{CastroNeto2009}, as they should. 

Let us now apply the HHK method to graphene systems.
We begin showing results of the LDOS for bulk graphene in the absence of external magnetic fields. 
Throughout this paper we use the tight-binding parameters given in Table \ref{tab:Parameterization}.

The optimal number of iterations $M$ depends on $P$, the size of the system chosen to represent the bulk, as well as on the desired accuracy. 
A detailed analysis about the dependence of the HHK method accuracy on $M$ for continuum spectra can be found in Ref.~\cite{Haydock1975}.  
For graphene in the absence of magnetic field, we find that $M\approx \sqrt{P}$ guarantees a good accuracy. 
At the end of this section we study the accuracy for the case where $B \neq 0$.

\begin{table}[h]
    \centering
    \caption{Tight-binding hopping parameter sets in eV.}
    \label{tab:Parameterization}
\begin{ruledtabular}
\begin{tabular}{c c c c}
    Parameterization & $t^{(1)}$~(eV) & $t^{(2)}$~(eV) & $t^{(3)}$~(eV) \\
    \hline
    A \cite{CastroNeto2009} & -2.7 &  &  \\
    B \cite{Charlier1991} & -3.0 & 0.3 &  \\
    C \cite{Lherbier2012} & -3.0933 & 0.19915 & -0.16214\\
    \end{tabular}
\end{ruledtabular}
\end{table}

\begin{figure}[h]
\centering
\includegraphics[width=0.9\columnwidth]{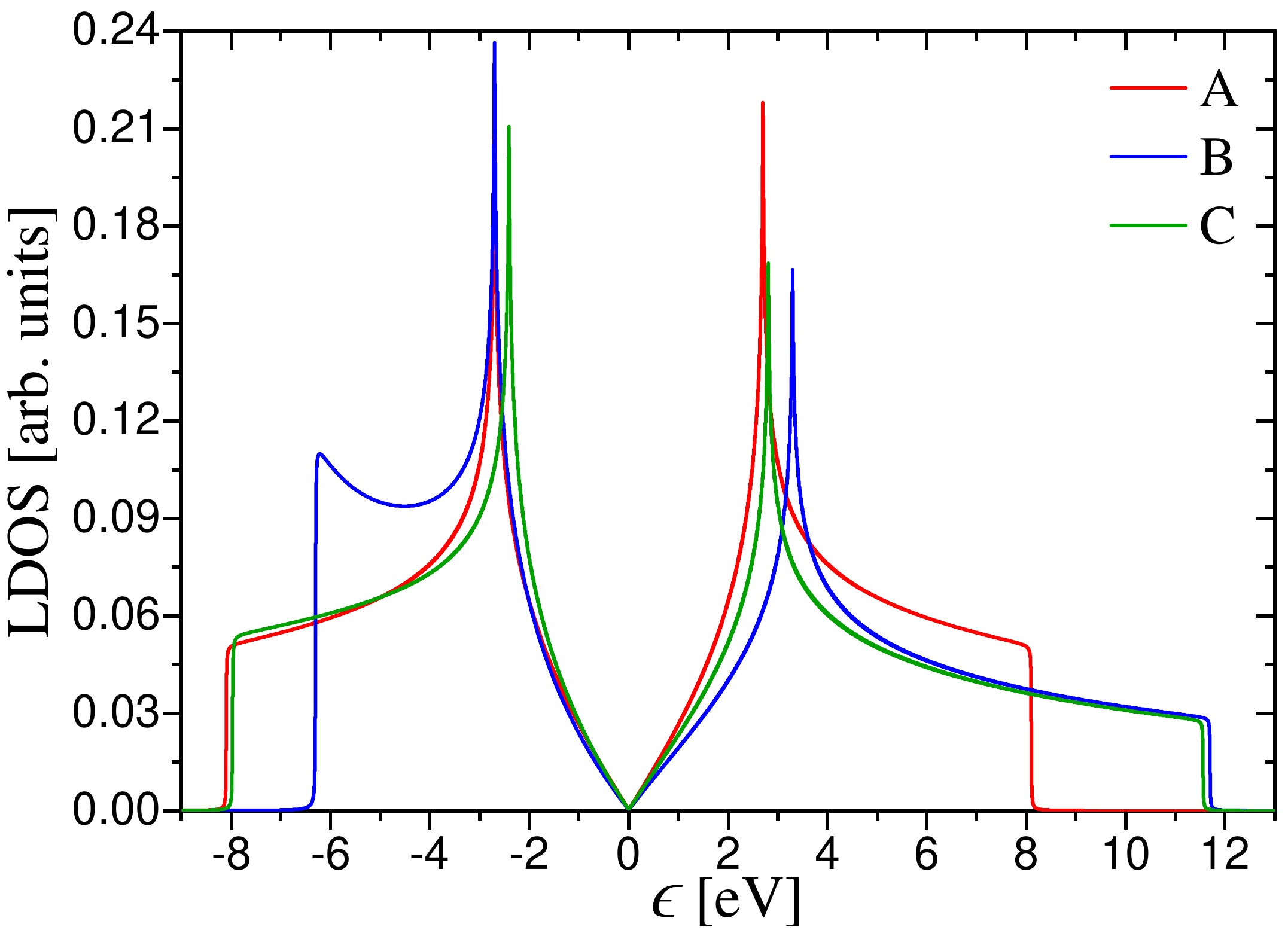}
\caption{Graphene LDOS (in arbitrary units)  as a function of the energy $\epsilon$ (in eV) calculated with the HHK method for the tight-binding Hamiltonian parameterizations A, B, and C. 
We set $\epsilon_0 = 0$ for better visualization.}
\label{fig:LDOS_graphene}
\end{figure}

Figure \ref{fig:LDOS_graphene} shows the LDOS of a graphene monolayer obtained from the HHK method for $P=2.5\times 10^{7}$ carbon atoms and $M=5000$ iterations. 
We have contrasted the the latter with $\rho (\epsilon)$ computed by a direct numerical evaluation of $\rho(\epsilon) = \frac{1}{{\cal A}_{\rm BZ}} \sum_{\textbf{k}}\delta(\epsilon -\epsilon_{\textbf{k}\lambda})$, where $\epsilon_{\textbf{k}\lambda}$ is given by Eq.~\eqref{eq:Energy dispersions} and the wavevectors ${\bf k}$ are sampled over the Brillouin zone of area ${\cal A}_{\rm BZ}$. 
By accounting for the normalization factor that relates LDOS$(j, \epsilon)$ with $\rho (\epsilon)$, we observe a nice agreement within the numerical precision imposed by the regularization parameter $\eta$. 
We find that differences are only appreciable, as expected, at the band edges as well as at the van Hove singularities. 

\begin{figure}[h!]
\includegraphics[width=0.9\columnwidth]{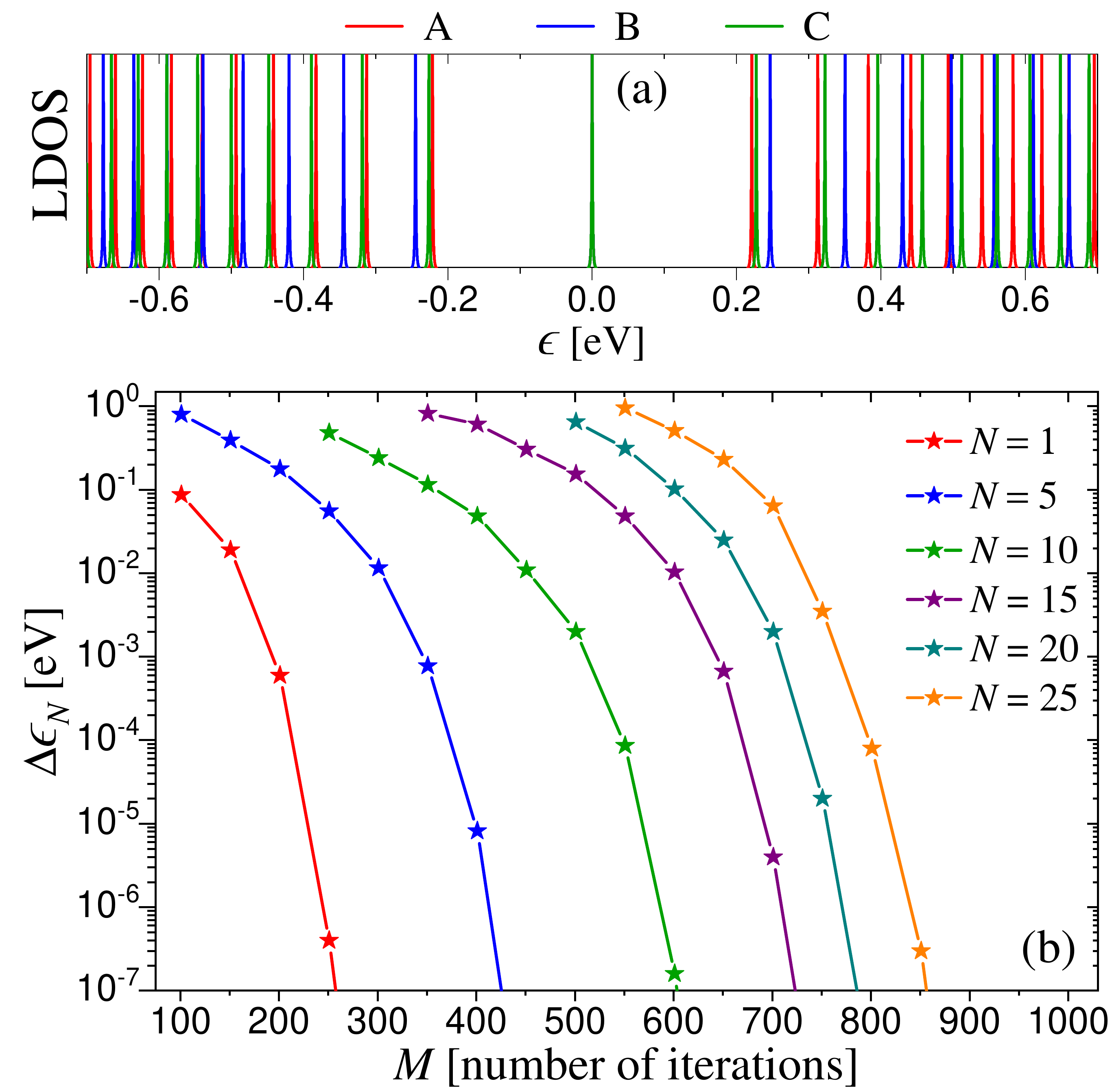}
\caption{(a) LDOS (in arbitrary units) versus $\epsilon$ (in eV) for $B = 25$~T, $\eta= 0.1$~meV and $M^*=1500$ for the tight-binding parameterizations A (red), B (blue) and C (green line). 
(b) Accuracy of the LL energies $\Delta \epsilon_{N} = |\epsilon_{N}^{M} - \epsilon_{N}^{M^*}|$ as a function of 
the number of iterations $M$ for the tight-binding parameterization A. 
Here, $\epsilon_{N}^{M^*}$ stands for the energy of the $N$th Landau level converged to a precision better than $\delta = 10^{-8}$~eV.}
\label{fig:LLs_iterations}
\end{figure}

Let us now consider the case of a pristine graphene sheet under an external perpendicular magnetic field $B$. 
The electronic spectrum becomes discrete and strongly degenerate forming a sequence of Landau levels. 
Figure~\ref{fig:LLs_iterations}(a) shows the graphene LDOS for the tight-binding parameterizations A, B, and C computed with the HHK method for $B=25$~T, $P=2.25 \times 10^6$, $M=1500$, and $\eta = 0.1$~meV. 
Due to the finite $\eta$, the LL are broadened and become Lorentzian distributions, whose energy peaks (obtained by fitting) are associated with the LL energies $\epsilon_N$.
Here, we chose $\eta \ll \vert \epsilon_{N} - \epsilon_{N-1} \vert$, guaranteeing that the $|N| \le 30$ lowest LL peaks are nicely resolved.

Figure \ref{fig:LLs_iterations}(b) shows the convergence of the Landau level energies $\epsilon_{N}$ as a function of $M$ for a lattice size of $P=2.25\times 10^{6}$, $B=25$~T, $\eta = 0.1$~meV and the tight-binding parameterization A.
Throughout this paper we set the absolute numerical precision to $\delta = 10^{-8}$~eV. 
We find that by setting $M^*=1500$, the accuracy of $\epsilon_N$ is better than $\delta$ for $|N| \le 30$.
Figure~\ref{fig:LLs_iterations}(b) shows that the LL energies $\ve_N$ converge rapidly to their asymptotic values as the number of iterations $M$ increases. 
The results indicate that  the number of iterations $M$ necessary to obtain a given precision $\delta$ scales as $M \sim \sqrt{|N|}$.

Let us now benchmark the HHK results against another well-established method to compute the LL energies \cite{Lado2016}.
The most straightforward numerical solution solves the eigenvalue problem for supercells with periodic boundary conditions whose size are dictated by the magnetic field strength and the phases of the Peierls substitution \cite{Brown1964}. 
Here, the computation of the LL spectra involves the diagonalization of matrices of dimension $\sim 9.07 \times 10^4/B [{\rm T}]$ \cite{Pereira2011},
that is computationally very costly for realistic values of the magnetic field.
Alternatively, Ref.~\cite{Lado2016} proposed to consider systems with an infinite ribbon geometry where, by a suitable gauge choice, the magnetic field is encoded in the hopping terms transversal to the ribbon preserving a ``free" dynamics in the longitudinal direction. In this way, the primitive unit cell has $N_W$ sites that depend linearly on the ribbon width $W$.
The $k$-space domains where the bands are flat, corresponding to energies $\epsilon_{N}^{\rm GNR}$, are associated with bulk Landau levels, namely, $\epsilon_N \approx \epsilon_{N}^{\rm GNR}$. 
The accuracy relies on $W/\ell_B \gg 1$, where $\ell_B = \sqrt{\hbar/eB} \approx 26 \sqrt{B[{\rm T}]}$~nm. 

\begin{figure}[h!]
\includegraphics[width=0.9\columnwidth]{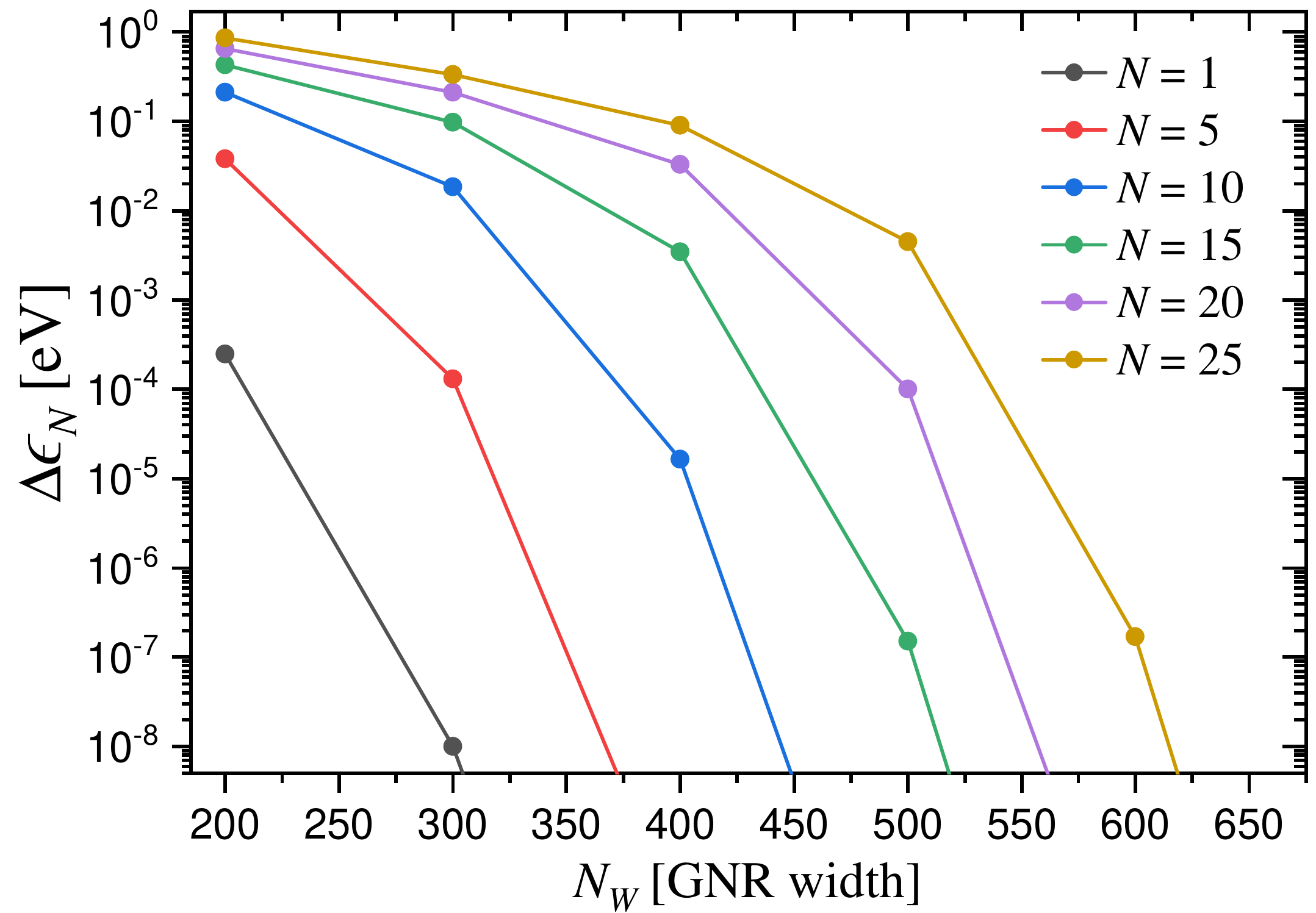}
\caption{Accuracy of the LLs energies $\Delta \epsilon_N = |\epsilon_N^{\rm GRN} - \epsilon_N^{M^*}|$, for graphene nanoribbons with zigzag edges as a function of their width $W = (3N_W/4  - 1) a_0$. }
\label{fig:LLs_GNR}
\end{figure}

We consider graphene nanoribbons with zigzag edges for which $W = (3N_W/4  - 1) a_0$. 
Figure~\ref{fig:LLs_GNR} shows the convergence of $\epsilon_N$ as a function of $N_W$. 
In this approach, an accurate assessment of $\epsilon_{N}^{\rm GNR}$, with negligible inter edge hybridization, requires increasing $W$ (or $N_{W}$) for larger $|N|$, with a computational cost that scales with $N_{W}^{3}$.
We note that by finding an optimized truncated basis set with $N_{\rm opt} < N_{W}$ one can significantly reduce computational time, but the number of operations of such procedure still scales as $N_{\rm opt}^3$.
Thus, by comparing Figs. \ref{fig:LLs_iterations}(b) and \ref{fig:LLs_GNR} and recalling that the HHK as an ${\cal O}(M)$ method, one concludes that the HHK method is rather unexpensive.

\subsection{Disorder}
\label{Disorder}

Several studies have addressed the effects of disorder in graphene in the quantum Hall regime \cite{Shon1998, Peres2006, Dora2007, Ostrovsky2008, Zhu2009}. 
Here, our goal is to obtain insight on disorder effects in the density of states (DOS) of large $\vert N\vert$ Landau subbands, particularly on their energy peaks, rather than studying specifics of a given experimental setup. 
For this reason we consider a simple model that captures the main features of disorder in graphene, namely, the Anderson disorder model. 
The on-site energy $u_{i}$ is assumed to be uniformly distributed between $-U$ and $+U$, that is, with zero mean $\left\langle u_{i} \right\rangle =0$ and variance $\langle\langle u_{i}^{2} \rangle\rangle = U^{2}/3$. 
This model describes the short range disorder regime \cite{Shon1998, Dora2007, Ostrovsky2008} where the impurities scattering range is much smaller than the lattice constant and, thus, impurity scattering mixes the $\xi=1$ (or $K$) and the $\xi=-1$ (or $K'$) valley spinors. 

We compare our numerical simulation with analytical results. In a pioneering work, Shon and Ando \cite{Shon1998} used the self-consistent Born approximation (SCBA) to calculate self-energy impurity average $\Sigma$ in graphene for different kinds of disorder. 
For short range impurities \cite{Shon1998},
\begin{align}
\label{eq:Self-energy_Ando}
\Sigma(\epsilon)=\dfrac{\gamma^{2}}{4} \sum^{N_{c}}_{N=-N_{c}} \dfrac{1}{\epsilon-\epsilon_{N}-\Sigma(\epsilon)},
\end{align}
where
\begin{align}
\gamma^{2}=\dfrac{\langle\langle u_{i}^{2} \rangle\rangle n_{\rm imp}}{\pi \ell_{B}^{2}}
\end{align}
is the parameter characterizing the disorder scattering strength and $n_{\rm imp}$ denotes the impurity concentration. 
The cutoff $N_{c}$ is introduced, to account for the finite bandwidth.

Contrary to the constant spacing of LL in two-dimensional electron gas systems, in graphene the spacing decreases with $\hslash\omega_{c}/\sqrt{\vert N\vert}$. 
This feature makes the evaluation of $\Sigma(\epsilon)$ somewhat more involved than in the standard case \cite{Ostrovsky2008}.
In the limit of weak scattering strength $\gamma^{2} \ll 1$ and considering the usual inequalities $\vert \epsilon -\epsilon_{N}\vert$, $\vert \Sigma\vert \ll \epsilon_{N}$ and $\vert \epsilon -\Sigma\vert < \hslash\omega_{c}/\sqrt{\vert N\vert}$ that correspond to  isolated LL subbands, the density of states is given by \cite{Ostrovsky2008}
\begin{align}
\label{eq:DOS_Ostrovsky}
{\rm DOS}(\epsilon)=\dfrac{2}{\pi\gamma^{2}} \sqrt{\gamma^{2} \left( 1- \chi\right) - \left[ \epsilon - \epsilon_{N}\left( 1-\chi\right)\right]^{2}},
\end{align}
where
\begin{align}
\chi = \dfrac{\gamma^{2}}{\left( \hslash\omega_{c}\right)^{2}} \ln{\left( D/\epsilon_{N}\right)},
\end{align}
and $D = \hbar\omega_{c}\sqrt{N_{c}}$ is a Debye-like energy cutoff. 
The DOS is a sequence of semi-circles (or better, elliptical shapes) with maxima at $\epsilon_{N}\left( 1-\chi\right)$ and full width at half maximum $\Gamma_{N}=\gamma\sqrt{3\left( 1-\chi\right)}$. 
Hence, disorder induces a $\Delta_{N}=-\chi\epsilon_{N}$ shift in the LL subband peaks.  
Note that by taking $\chi = 0$, one recovers the simple single-isolated subband approximation found in Ref.~\cite{Shon1998}.


\section{Results}
\label{sec:results}

In this section we use the HHK method to study the LL energies $\epsilon_{N}$, as a function of $\vert N\vert$, the external magnetic field $B$, and disorder. We also compare the results with the experimental data \cite{Plochocka2008}.

\subsection{Landau levels in pristine graphene}
\label{sec:pristineLL}

The analysis of the LL spectra measured in Ref.~\cite{Plochocka2008} reports a discrepancy between the analytical approximation for $\epsilon_N$ and the corresponding experimental values. The problem has been attributed to the model Hamiltonian truncation to second nearest neighbor hopping terms.  
Here, we expand the original theoretical analysis by accounting for third nearest neighbor terms and compare our results both with numerical calculations and with the experimental data. 

The dispersion relation of graphene can be expressed as a power series in $\bf q$, defined as ${\bf q} = {\bf k} \mp {\bf K}$, where $\pm {\bf K}$ correspond to $\gamma_{\pm\textbf{K}} =\gamma_{\pm\textbf{K}}'=0$ that defines the so-called Dirac points. 
The external magnetic field is accounted for by minimal substitution, ${\bm \Pi} = \hbar {\bf q} + e{\bf A}$.

By using canonical quantization and the approximations suggested in Ref.~\cite{Goerbig2011} (see 
App.~\ref{sec:appendix} for details), we write the energies of the LLs in the large $\vert N\vert$ limit up to third-nearest neighbors hopping as
\be 
\label{eq:grapheneLL_analytic}
\widetilde{\epsilon}_{N} = \widetilde{\epsilon}_{N}^{(1)} + \widetilde{\epsilon}_{N}^{(2)} + \widetilde{\epsilon}_{N}^{(3)},
\ee
where the first two terms 
\begin{eqnarray}
\label{eq:grapheneLL_analytic_1}
\widetilde{\epsilon}_{N}^{(1)} & \approx & {\rm sgn}(N)\hslash\omega_{c} 
\vert N \vert^{1/2} \left(   1 - \dfrac{3}{8}  \dfrac{a_{0}^{2}}{\ell_B^{2}} \vert N \vert \right) ,
\\
\label{eq:grapheneLL_analytic_2}
\widetilde{\epsilon}_{N}^{(2)} & \approx & \hslash\omega_{c} \dfrac{t^{(2)}}{\vert t^{(1)}\vert }\dfrac{3}{\sqrt{2}}\dfrac{a_{0}}{\ell_{B}} \vert N \vert \left(   1 - \dfrac{3}{4}  \dfrac{a_{0}^{2}}{\ell_B^{2}} \vert N \vert \right) ,
\end{eqnarray}
were obtained in Refs. \cite{Plochocka2008,Goerbig2011}. Our derivation expands the latter results to account for $t^{(3)}$, namely
\be
\label{eq:grapheneLL_analytic_3}
\widetilde{\epsilon}_{N}^{(3)} \approx -{\rm sgn}(N)\hslash\omega_{c} \dfrac{2t^{(3)}}{t^{(1)}} \vert N \vert^{1/2} \left( 1 - \dfrac{t^{(3)}}{t^{(1)}} - \dfrac{59}{32}  \dfrac{a_{0}^{2}}{\ell_B^{2}} \vert N \vert \right) .
\ee
Note the subleading terms in the above expansions give minor corrections, since $a_0/\ell_B = 5.4 \times 10^{-3} \sqrt{B [{\rm T}]}$ is small for $B$ fields within the experimental reach.  

In what follows we show that Eqs.~\eqref{eq:grapheneLL_analytic_1} to \eqref{eq:grapheneLL_analytic_3} give an excellent approximation to the LLs energies obtained from the tight-binding Hamiltonian of Eqs.~\eqref{eq:H_tb} and \eqref{eq:Peierls}.

\begin{figure}[h!]
\includegraphics[width=0.9\columnwidth]{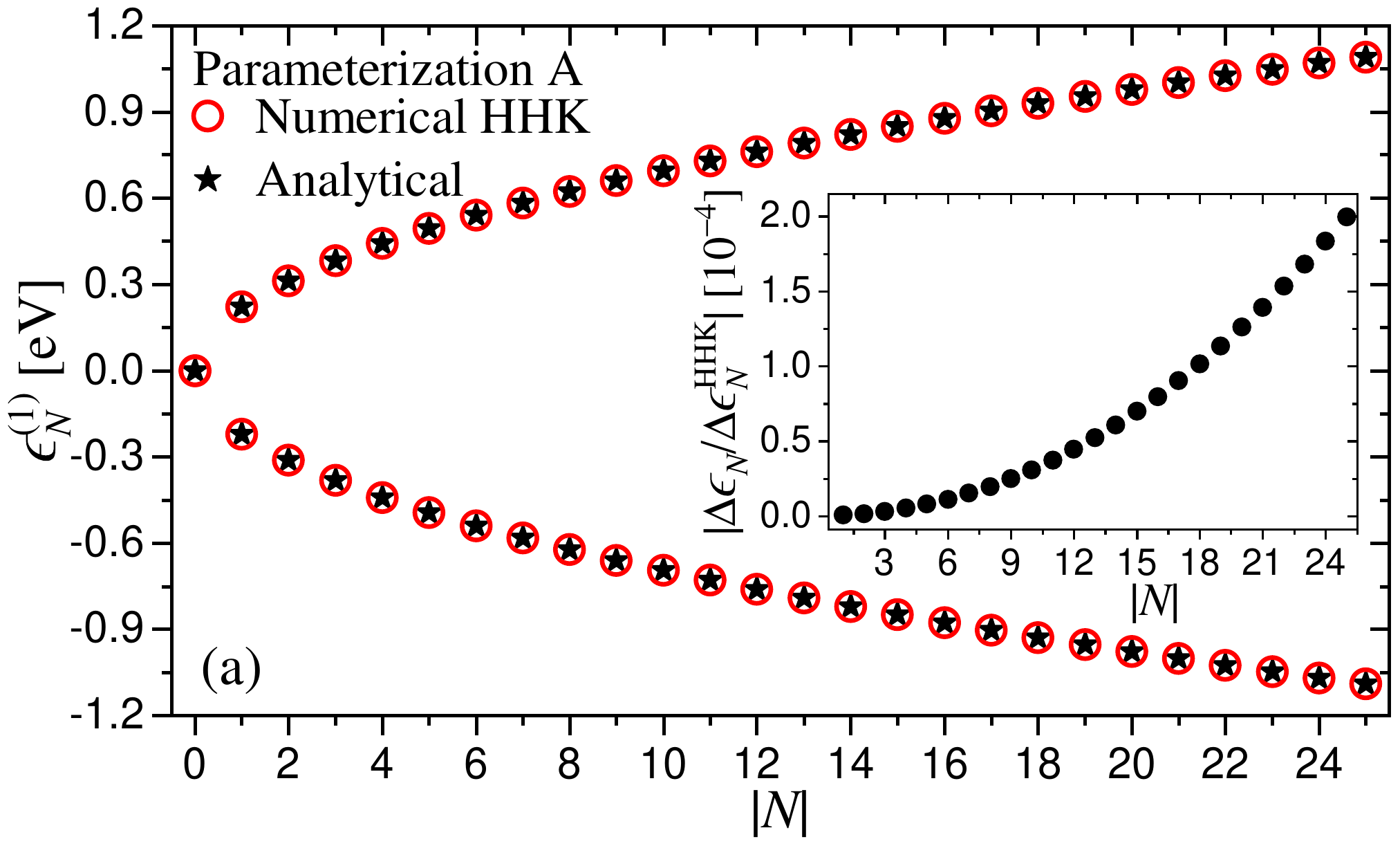}
\includegraphics[width=0.9\columnwidth]{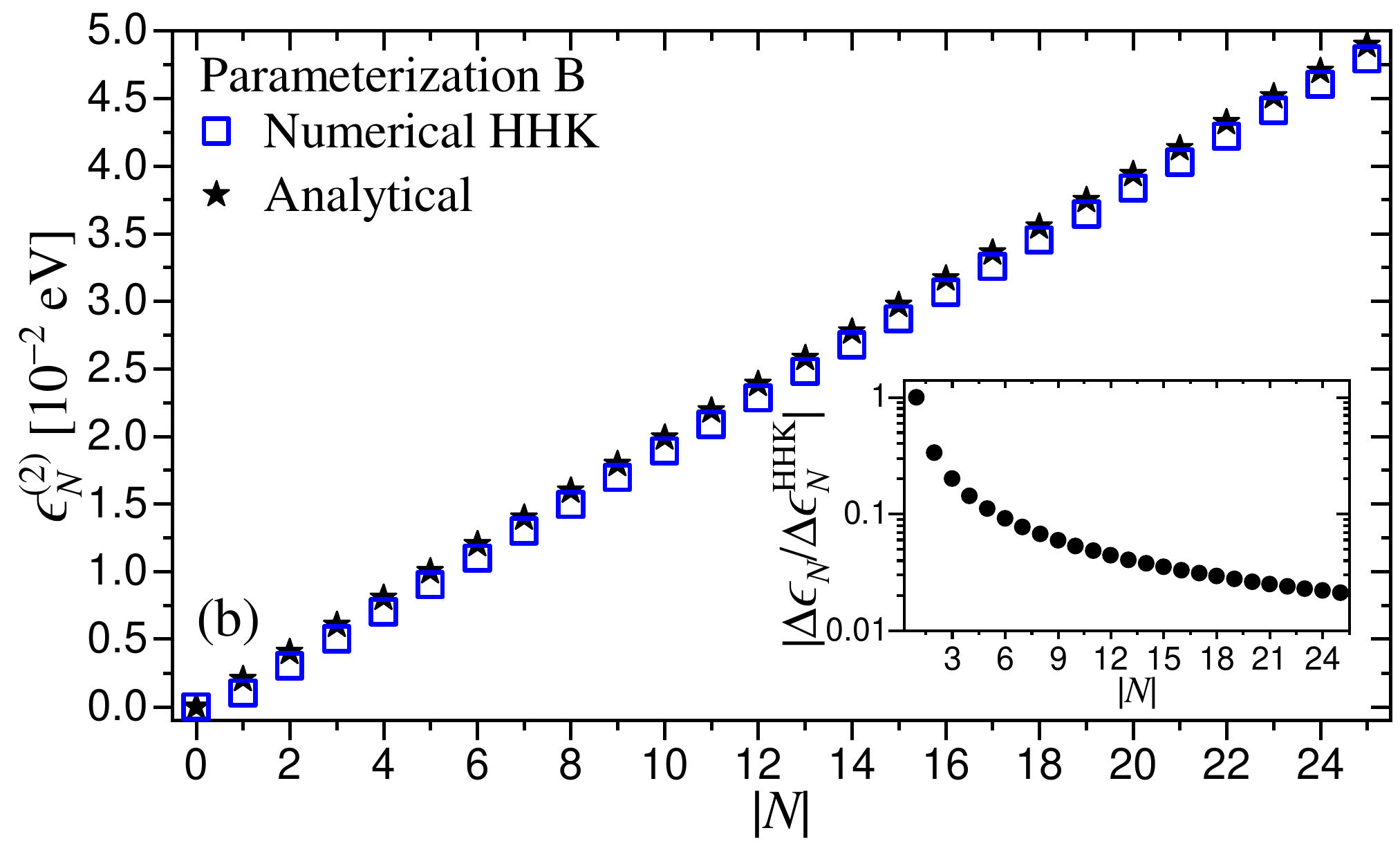}
\includegraphics[width=0.9\columnwidth]{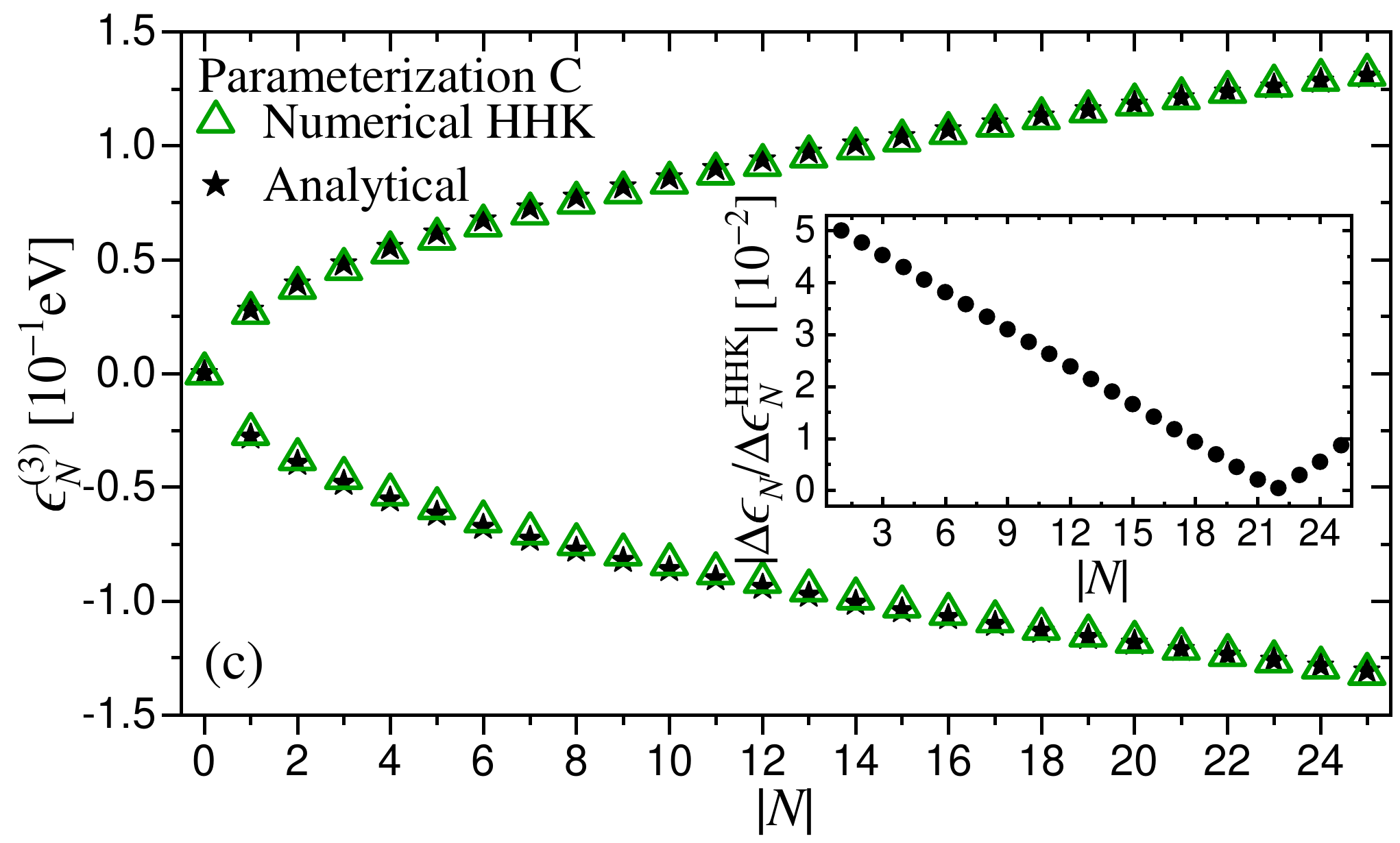}
\caption{Comparison between the analytical contributions to $\widetilde{\epsilon}_N$ and HHK results for a pristine graphene monolayer with $B=25$~T.
(a) first neighbor contribution, (b) second neighbor contribution and (c) third neighbor contribution.
The insets show the deviation between the HHK results and the analytical approximation, $\vert\Delta\epsilon_{N} /\epsilon_{N}^{{\rm HHK}}\vert$, see text for details.}
\label{fig:LLs_corrections}
\end{figure}

Figure~\ref{fig:LLs_corrections}(a) compares the analytical $\widetilde{\epsilon}_N^{(1)}$ with the numerical HHK results $\epsilon_N^{(1)} = \epsilon_N^{\rm HHK}$ for the  first nearest neighbor tight-binding Hamiltonian, namely, the parameterization A. 
The insert shows the relative deviation $|\Delta \epsilon_N /\epsilon_N^{\rm HHK}|$ versus $N$, where $\Delta \epsilon_N = \epsilon_N^{\rm HHK} -\widetilde{\epsilon}_N^{(1)}$. 
The agreement is very good with a relative deviation smaller than  $2.3 \times 10^{-4}$ for $|N| < 25$. 
The results indicate that $\Delta \epsilon_N \propto |N|^{5/2}$, a scaling corresponding to terms related to $q^3$ that have been neglected in the momentum expansion, see App.~\ref{sec:appendix}.

Figure~\ref{fig:LLs_corrections}(b) contrasts the analytical $\widetilde{\epsilon}_N^{(2)}$ second nearest neighbor contribution to the LL spectrum with its numerical counterpart $\epsilon_{N}^{(2)}$.
The later is defined as $\epsilon_{N}^{(2)} \equiv \epsilon_{N} - \epsilon_{N}^{(1)}$, where $\epsilon_{N}$ and 
$\epsilon_{N}^{(1)}$ are computed numerically. 
Both are obtained using the tight-binding parameter set B, but for the calculation of $\epsilon_{N}^{(1)}$ we set $t^{(2)}=0$ to single-out the second-nearest neighbor contribution. 
The insert shows $|\Delta \epsilon_N /\epsilon_N^{\rm HHK}|$ versus $N$, $\Delta \epsilon_N = \epsilon_N^{\rm HHK} -\widetilde{\epsilon}_N^{(1)}$.
Here $\Delta\epsilon_{N}$ is approximately $\sim 1.01\times 10^{-3}$~eV for $\vert N \vert \lesssim 25$. 
The origin of $\Delta\epsilon^{(2)}$ is small discrepancy between the constant diagonal term that shifts the energy spectrum by $-3t^{(2)}$ ($t^{(2)}=0.3$~eV, see Table \ref{tab:Parameterization}) in the analytical approach and the HHK results where the shift is $-0.899$~eV.

Finally, Fig.~\ref{fig:LLs_corrections} (c) compares $\widetilde{\epsilon}_N^{(3)}$ with $\epsilon^{(3)}$. 
Here, we define $\epsilon_{N}^{(3)}=\epsilon_{N}-\epsilon_{N}^{(1)}-\epsilon_{N}^{(2)}$, where we use the HHK method to compute $\epsilon_{N}$ with the parameter set C and $\epsilon_{N}^{(1)}+\epsilon_{N}^{(2)}$ by taking $t^{(3)}=0$.
The results indicate that the third nearest neighbor contributions to $\epsilon_{N}$ are nicely described by the leading $\vert N\vert ^{1/2}$ term.
Here, in distinction to the first-nearest neighbors, the deviation is $\Delta\epsilon\sim \vert N \vert ^{3/2}$ but still small, $|\Delta \epsilon_N /\epsilon_N^{\rm HHK}| \leq 5\times 10^{-2}$ for $\vert N \vert \leq 25$.

\begin{figure}[h!]
\includegraphics[width=1.0\columnwidth]{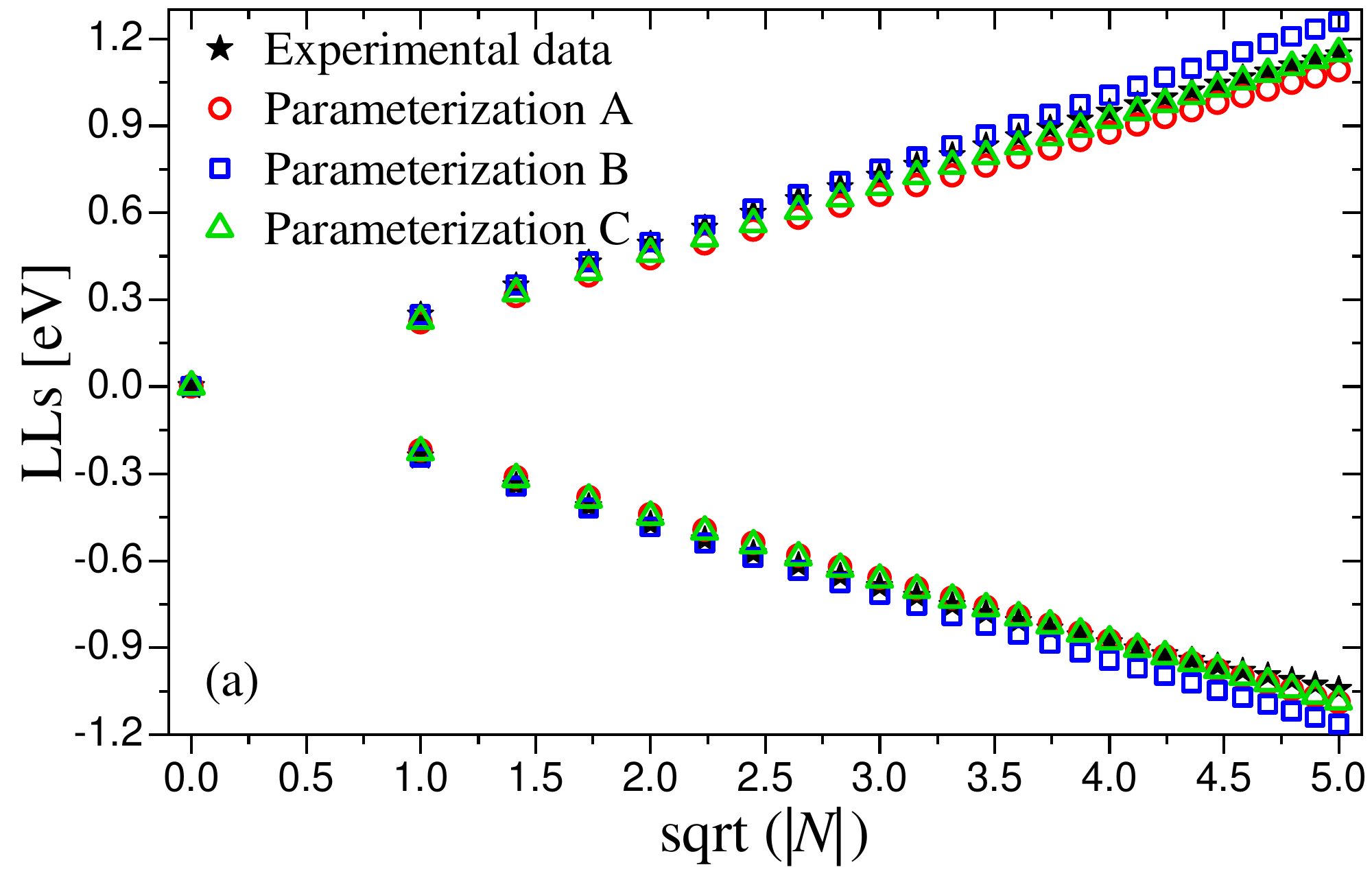}
\includegraphics[width=1.0\columnwidth]{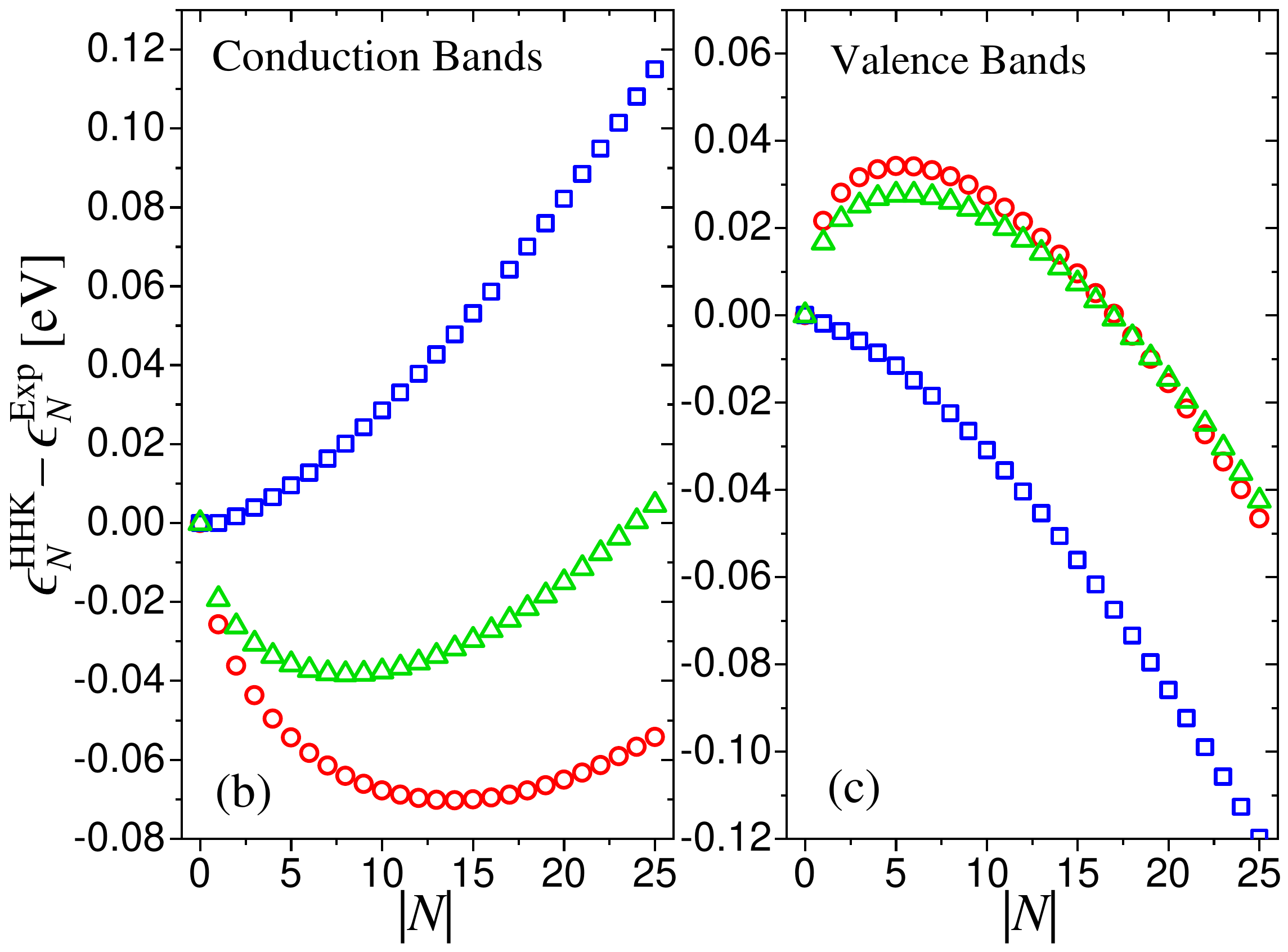}
\caption{(a) The Landau levels spectrum as a function of $\sqrt{\vert N\vert}$ for the tight-binding parameterizations A, B and C (see Table \ref{tab:Parameterization}) and $B=25$~T.
The experimental data are obtained from Ref.~\cite{Plochocka2008}. 
Deviation between the computed and the experimental LL energies, $\epsilon_{N}^{\rm HHK} - \epsilon_{N}^{\rm exp}$, for the (b) conduction and (c) valence bands.}
\label{fig:HHK&experimental}
\end{figure}

Let us now compare our results with the experimental data\cite{Plochocka2008}. Figure~\ref{fig:HHK&experimental} (a) shows that, since to leading order $\epsilon_{N}\sim \vert N \vert ^{1/2}$, all tight-binding parameterizations correctly capture the main trend of $\epsilon_{N}$. 
In Fig.~\ref{fig:HHK&experimental} (b) and \ref{fig:HHK&experimental} (c) we show $\epsilon_{N}^{\rm HHK}-\epsilon_{N}^{\rm exp}$ for the conduction and the valence bands.
We find that the parameterization B (up to second nearest neighbor hopping), considered in Ref. \cite{Plochocka2008} shows the most agreement with the experimental data, with $\vert \epsilon_{N}^{\rm HHK}-\epsilon_{N}^{\rm exp}\vert \propto \vert N \vert ^{3/2}$.
This explains the effectiveness on the extra arbitrary ``\scriptsize$\mathcal{W}$\normalsize" parameter used in Ref.\cite{Plochocka2008,Goerbig2011} to obtain a good fit.
The parameterization C (up to third nearest neighbor hopping) gives the best results, but still shows some deviations from the experiments.
We stress that the tight-binding parameters are fits of DFT calculations to $\epsilon_{\bf k}$ for $B=0$ and not to the LL energies $\epsilon_{N}$.

\subsection{Disordered graphene}

Let us now investigate the effect of disorder in the LL spectrum of bulk graphene. 
We briefly address the disorder induced LL broadening $\Gamma_{N}$ but our main interest is the energy shift $\Delta_{N}$.

Disorder breaks translational invariance and the sites are no longer equivalent. 
By involving the ergodic hypothesis one identifies the ensemble average of the LDOS at a given system site with the DOS of the disordered system. 
In this study, the ensemble averages involve $10^{4}$ disorder realizations. 
We use $\eta =0.5$~meV as regularization parameter. 
This choice guarantees that $\eta \ll \Gamma_{N}$ for the disorder strengths we study and, thus, has a negligible effect on the results.

Since the short-range disorder effects on the DOS are not expected to depend on the tight-binding parameters, here we only consider the parameterization A.
The SCBA equation, Eq. (\ref{eq:Self-energy_Ando}), can be numerically solved by a simple iteration starting with a given initial value of the self-energy. 
We have introduced a Debye-like cutoff $N_{c}=1342$ and start the self-consistent loop with an initial self-energy of ${\rm Im}\; \Sigma (\epsilon) = 1$~meV. 
The iteration converges very rapidly.

\begin{figure}[h!]
\includegraphics[width=1\columnwidth]{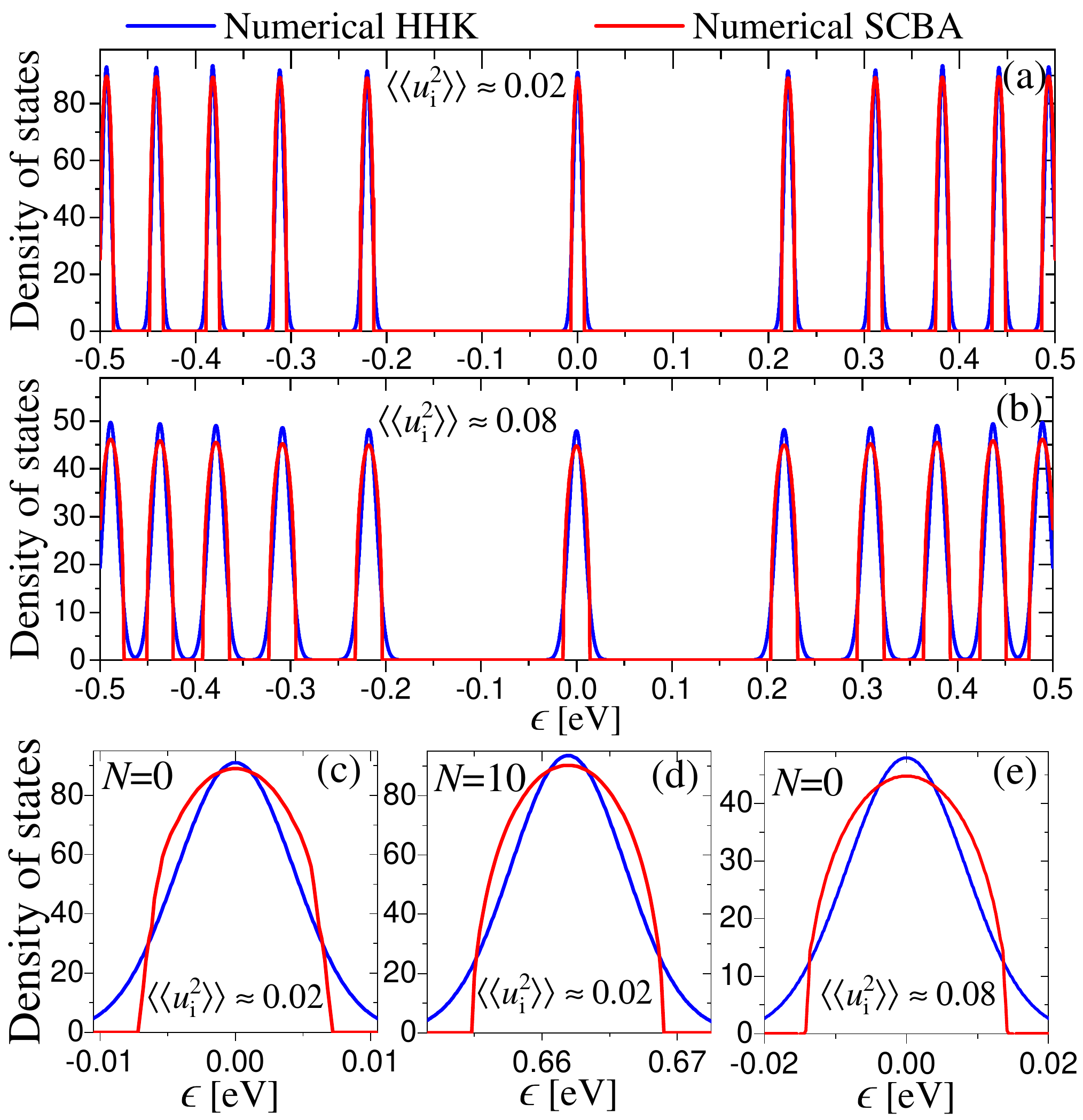}
\caption{DOS (in arbitrary units) versus $\epsilon$ (in eV) of disordered graphene monolayers for the tight-binding parameterization A and $B=25$~T obtained by the HHK method and the SCBA for (a) $\langle\langle u_{i}^{2} \rangle\rangle \approx 0.02$ and (b) $0.08$. 
Blow up of the subband DOS for (c) $\langle\langle u_{i}^{2} \rangle\rangle \approx 0.02$, $N=0$; (d) $\langle\langle u_{i}^{2} \rangle\rangle \approx 0.02$, $N=10$; and (e) $\langle\langle u_{i}^{2} \rangle\rangle \approx 0.08$, $N=0$.}
\label{fig:DOS_HHS_vs_SCBA}
\end{figure}

Figure~\ref{fig:DOS_HHS_vs_SCBA} compares the HHK with the SCBA results for the DOS($\epsilon$) of graphene at $B=25$~T for different disorder strengths, namely, $\langle\langle u_{i}^{2} \rangle\rangle \approx 0.02$ and $0.08$. 
The Landau subbands are broadened and shifted toward the zero-energy in the presence of on-site disorder with the exception of the zero-energy subband which is only broadened, as nicely shown by Figs.~\ref{fig:DOS_HHS_vs_SCBA}(a) and \ref{fig:DOS_HHS_vs_SCBA}(b).

Figures~\ref{fig:DOS_HHS_vs_SCBA}(c) to \ref{fig:DOS_HHS_vs_SCBA}(e) compare the HHK with the SCBA density of states of different LL subbands.
The SCBA predicts a semi-circular (or, more precisely, elliptical) shaped DOS$(\epsilon)$ for all $N$.
For IQH systems with quadratic dispersion, it is well established that the DOS of the $N=1$ subband has Gaussian-like shape\cite{Wegner1983} and the semi-circular DOS is expected for $N\gg 1$.
Our results show large deviations from the SCBA independent of $N$.
This observation is consistent with previous numerical investigations of the DOS in graphene at the QH regime\cite{Zhu2009} and deserve further investigation.

\begin{figure}[h!]
\includegraphics[width=0.8\columnwidth]{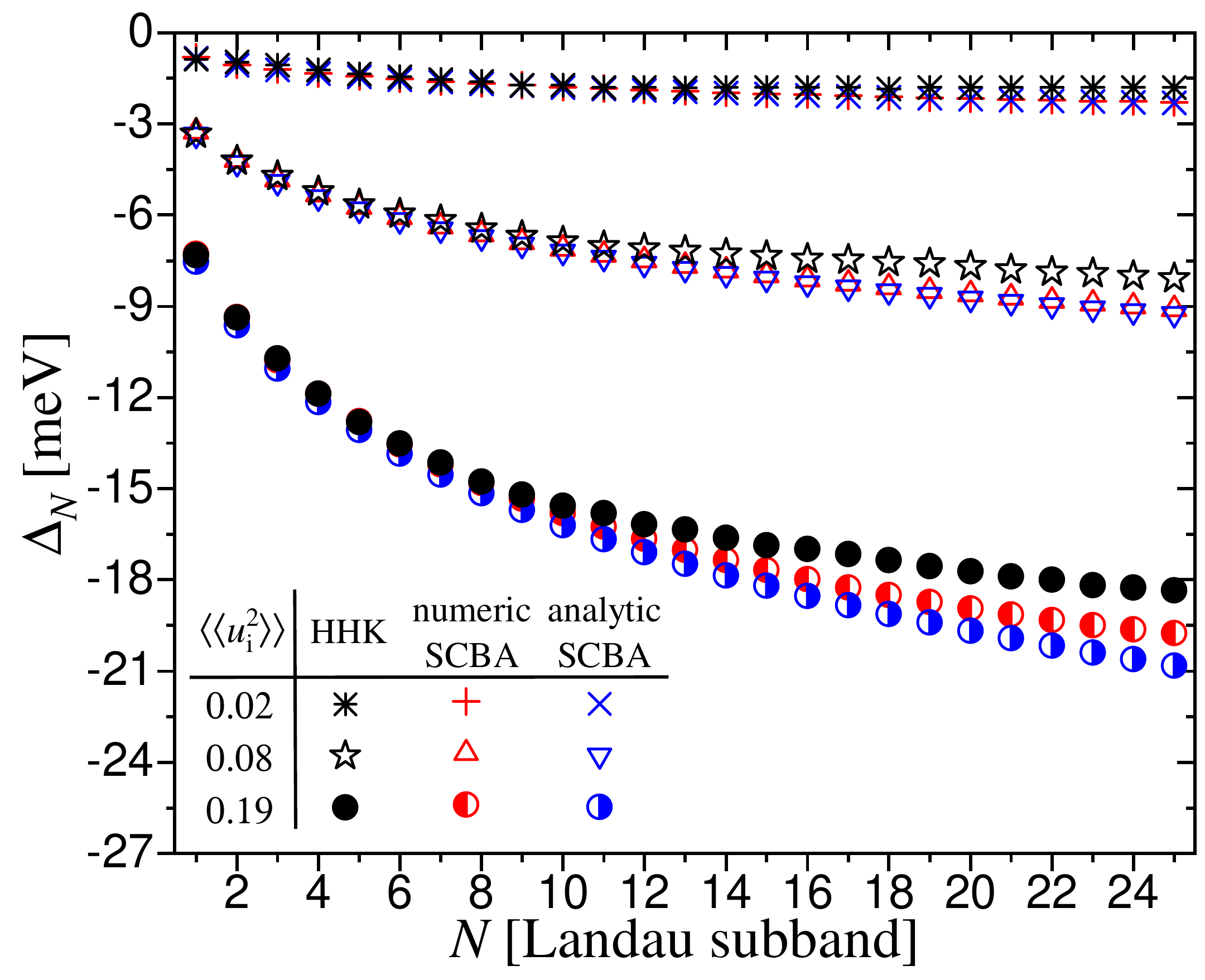}
\caption{Renormalization of the LL subband peak energy $\Delta_{N}$ as function of $N$ for $\langle\langle u_{i}^{2} \rangle\rangle \approx 0.02$, $0.08$ and $0.19$. The black and red symbols correspond to the numerical results of HHK and SCBA, respectively, and the blue symbol corresponds to the analytical prediction of Ref. \cite{Ostrovsky2008}.}
\label{fig:Delta_N}
\end{figure}

Figure \ref{fig:Delta_N} shows the disorder renormalization of the Landau level subbands peak energies $\Delta_{N}$.
We define $\Delta_{N}^{\rm HHK}$ as the difference between the ensemble average $\langle \epsilon_{N}^{\rm HHK} \rangle$ and the pristine value $\epsilon_{N}^{\rm HHK}$.
The $\Delta_{N}^{\rm SCBA}$ can be obtained from the numerical solution of Eq. (\ref{eq:Self-energy_Ando}) or evaluated from the approximation given by Eq. (\ref{eq:DOS_Ostrovsky}).
The agreement is very good.

The results show that disorder contributes to increase the deviation between theory and experiment.
However, for realistic disorder strengths, $\Delta_{N}$ is very small.

\section{Conclusions and discussion}
\label{sec:conclusion}

In this paper we have studied the large $\vert N \vert$ Landau level spectrum of graphene monolayers beyond the first-nearest neighbor tight-binding approximation and the effective low-energy Dirac Hamiltonian.
The study has to goals, namely, to discuss the effect of different kinds of hopping processes to the high-energy LLs in graphene and to introduce an efficient numerical method for this analysis.

Regarding the methodology, our results show that the HHK method\cite{Haydock1972, Haydock1975, Haydock1980} is very efficient for the computation of the LDOS of graphene in the QH regime. 
This method has a long record of success, but to the best of our knowledge has not been used to study discrete spectra.
Our paper shows that the HHK method is very accurate and, since it is an $O({\cal N})$ approach, it is, computationally much faster then other methods found in the literature \cite{Lado2016,Pereira2011}.

As for the large-$\vert N \vert$ LL spectrum, we show that the analytical solution of the continuum (long wavelength) effective graphene Hamiltonian is very accurate up to $\vert N \vert \leq 25$ and $B=25$~T.
We find that by including third nearest hopping processes into the tight-binding model Hamiltonian the agreement between theory and experiment is significantly improved.
Although the tight-binding parameters can vary depending on the methodology employed to extract them from DFT calculations, the variations are small and we expect our results to be robust.

We also analyzed the effect of disorder.
We find a very good agreement between numerical simulations and the SCBA for the disorder induced renormalization of the LL subbands energy peaks $\Delta_{N}$.
For realistic disorder strengths, $\Delta_{N}$ gives a small correction to $\epsilon_{N}$ and, thus, does not impact our conclusions.
Interestingly, our simulations indicate that even for large-$\vert N \vert$ the subband DOS does not show a semi-circular shape, at odds with the theoretical studies of 2D electron gas systems in the QH regime \cite{Benedict1986,Carra1989,Efetov1989,Dmitriev2012}. 
We believe this issue deserves further investigation.

We expect the HHK method to be usefull for the analysis of the LDOS of other QH 2D systems and, since it does not rely on periodic boundary conditions, it can also used in the study of quasicrystals \cite{Ahn2018,Yao2018} and fractal lattices \cite{Kempkes2019}.

\acknowledgements
This work was partially supported by the Brazilian Institute of Science and Technology (INCT) in Carbon Nanomaterials and the Brazilian agencies CAPES, CNPq, FAPEMIG, and FAPERJ.

\appendix
\section{Landau levels in graphene: Analytical approach}
\label{sec:appendix}

As mentioned i the main text, the low energy spectrum of graphene is obtained by expanding the phase factors $\gamma$ and $\gamma '$ in powers of $\textbf{q}$ up to third order in $\textbf{q}$.
For the series expansion, it is convenient\cite{Bena2009} to use $\gamma\rightarrow\gamma\exp{\left( i{\bf k}\cdot\bm{\delta}_{3}\right)}$ and $\gamma '\rightarrow\gamma '\exp{\left( i2{\bf k}\cdot\bm{\delta}_{3}\right)}$ where $\bm{\delta}_{3}=-a_{0} \hat{\textbf{e}}_{y}$. 
These changes do not affect the dispersion relation and render more symmetric expressions to work with \cite{Bena2009}.
Thus, the expansions of the phase factors read
\begin{align}
\label{eq:gamma}
\gamma^{\xi}_{\bf q} \approx & - \xi \dfrac{3a_{0}}{2}\left( q_{x} + i \xi q_{y} \right)
\nonumber\\
&+ \dfrac{3a^{2}_{0}}{8}\left( q_{x} - i \xi q_{y} \right)^{2}
+ \xi \dfrac{3a^{3}_{0}}{16}\vert {\bf q}\vert^{2} \left( q_{x} + i \xi q_{y} \right) ,
\end{align}
\begin{align}
\label{eq:gammalinha}
\gamma '^{\xi}_{\bf q} \approx & - \xi 3a_{0}\left( q_{x} + i \xi q_{y} \right)
\nonumber\\
&+ \dfrac{3a^{2}_{0}}{2}\left( q_{x} - i \xi q_{y} \right)^{2}
+ \xi \dfrac{3a^{3}_{0}}{2}\vert {\bf q}\vert^{2} \left( q_{x} + i \xi q_{y} \right) ,
\end{align}
where $\xi=\pm 1$ denotes the valley index and $\vert {\bf q} \vert a_{0} \ll 1$ guarantees the accuracy of the approximation.

The energy dispersion relation in power of $\vert {\bf q} \vert a_{0}$ is obtained by combining the relations (\ref{eq:gamma}), (\ref{eq:gammalinha}) and (\ref{eq:Energy dispersions}).
As a result, the energy dispersion up to third-nearest neighbors hopping is conveniently written as
\be
\label{eq:energy dipersion without field}
\epsilon_{\lambda\xi\textbf{q}} = \epsilon_{\lambda\xi\textbf{q}}^{(1)} + \epsilon_{\xi\textbf{q}}^{(2)} + \epsilon_{\lambda\xi\textbf{q}}^{(3)} ,
\ee
where
\begin{eqnarray}
\epsilon_{\lambda\xi\textbf{q}}^{(1)} &\approx &\lambda\hslash v^{(1)}_{F} \vert {\bf q}\vert \left[ 1-\xi\dfrac{a_{0}\vert {\bf q}\vert}{4} \cos{\left( 3\varphi_{\bf q}\right) } \right] ,
\\
\epsilon_{\xi\textbf{q}}^{(2)} &\approx &\hslash v^{(1)}_{F} \dfrac{t^{(2)}}{\vert t^{(1)}\vert}\dfrac{3a_{0}}{2} \vert {\bf q}\vert^{2} \left[ 1-\xi\dfrac{a_{0}\vert {\bf q}\vert}{2} \cos{\left( 3\varphi_{\bf q}\right) } \right] , \;\;\;\;
\\
\epsilon_{\lambda\xi\textbf{q}}^{(3)} &\approx &\lambda\hslash v^{(1)}_{F} \dfrac{2t^{(3)}}{t^{(1)}} \vert {\bf q}\vert \bigg[ \cos{\left( 2\varphi_{\bf q}\right)}+\dfrac{t^{(3)}}{t^{(1)}}
\nonumber \\
&& -\xi a_{0}\vert {\bf q}\vert \left( \dfrac{3}{4}\cos{\left( \varphi_{\bf q}\right) } + \dfrac{t^{(3)}}{t^{(1)}}\cos{\left( 3\varphi_{\bf q}\right) } \right)\bigg] . \;\;\;\;
\end{eqnarray}
The term $\cos(3\varphi_{\textbf{q}})$ is remnant of the symmetry of the underlying lattice and gives rise to the so-called trigonal warping \cite{CastroNeto2009,Goerbig2011}. 
We note that the third nearest neighbors hopping terms modify the standard Fermi velocity $v^{(1)}_{F}=3a_{0}t^{(1)}/2\hslash$.

We include the magnetic field in the Dirac Hamiltonian by minimal substitution, that is, by replacing the canonical momentum $\textbf{p}=\hslash {\bf q}$ by the kinetic momentum, $\boldsymbol{\Pi}=\textbf{p}+e \textbf{A}(\textbf{r})$. 
The ansatz remains accurate as along as the lattice spacing is much smaller than the magnetic length, $\ell_{B}=\sqrt{\hslash /eB}$, a condition fulfilled even by the most intense magnetic field currently produced in laboratory \cite{Goerbig2011}.
It is convenient to express the operator $\boldsymbol{\Pi}$ in terms of harmonic-oscillator ladder operator namely \cite{Plochocka2008,Goerbig2011}
\be
\widehat{a}=\dfrac{\ell_{B}}{\sqrt{2}\hslash} \left( \Pi_{x} - i\Pi_{y} \right) \;\;\;{\rm and}\;\;\; \widehat{a}^{\dagger}=\dfrac{\ell_{B}}{\sqrt{2}\hslash} \left( \Pi_{x} + i\Pi_{y} \right)
\ee
with $\left[ \widehat{a}, \widehat{a}^{\dagger} \right] =1$. With the help of standard canonical quantization rules, one writes the Dirac Hamiltonian in the presence of an external magnetic field $\textbf{B}=B \hat{\rm \textbf{e}}_{z}$ as \cite{Plochocka2008}
\be
\label{Hamiltonian with field}
H_{B}^{\xi}\equiv \left( 
			\begin{array}{cc}
			h'_{\xi} & h_{\xi}^{\dagger} \\ 
			h_{\xi} & h'_{\xi} 
			\end{array} 
			\right),
\ee
where the diagonal element is
\begin{eqnarray}
\label{diagonal element with field}
h'_{\xi} &\thickapprox & \hslash \omega_{c}
\dfrac{t^{(2)}}{\vert t^{(1)}\vert }\dfrac{3}{\sqrt{2}}\dfrac{a_{0}}{\ell_{B}} \Bigg[ \widehat{a}^{\dagger}\widehat{a} - \xi\dfrac{1}{2\sqrt{2}}\dfrac{a_{0}}{\ell_{B}} \left( \widehat{a}^{\dagger 3}+\widehat{a}^{3} \right) 
\nonumber\\
&& \qquad -\>
\dfrac{3}{8} \dfrac{a_{0}^{2}}{\ell_{B}^{2}} \left( \widehat{a}^{\dagger}\widehat{a} \right)^{2} \Bigg]
\end{eqnarray}
and the off-diagonal ones are
\begin{eqnarray}
\label{off-diagonal element with field}
h_{\xi} & \thickapprox & \xi\hslash\omega_{c}  \Bigg[   \left( \widehat a^{\dagger}+\dfrac{2t^{(3)}}
{t^{(1)}}\widehat{a} \right) -\xi\dfrac{a_{0}}{\sqrt{2}\ell_{B}} \left( \dfrac{\widehat{a}^{2}}{2}+\dfrac{2t^{(3)}}{t^{(1)}} 
\widehat{a}^{\dagger 2} \right)
\nonumber\\
&& \qquad -\>
\dfrac{a_{0}^{2}}{\ell_{B}^{2}}\left( \dfrac{1}{4} \widehat{a}^{\dagger 2}\widehat{a}+\dfrac{2t^{(3)}}{t^{(1)}}\widehat{a}^{\dagger} \widehat{a}^{2}\right) \Bigg],
\end{eqnarray}
where $\omega_{c}=\sqrt{2}v^{(1)}_{F}/\ell_{B}$ is the cyclotron frequency.

Here, we are interested in the $\vert N\vert \gg 1$ limit. 
Thus, to a good approximation we can neglect $1/N$ corrections related to the ordering of the ladder operators\cite{Goerbig2011,Plochocka2008} and approximate $\left( \widehat a^{\dagger}\widehat a \right) ^{2} \approx \widehat a^{\dagger 2}\widehat a^{2} \approx \widehat a^{2} \widehat a^{\dagger 2}$.

Hence, solving the eigenvalue equation in two-spinors $H_{B}^{\xi}\psi_{N} = \epsilon_{N} \psi_{N}$, where $\psi_{N}= \left( u_{N},v_{N}\right)^T$, the eigenvalue equation for the second spinor component becomes \cite{Goerbig2011}
\be
\label{eq:secularequation}
h_{\xi}h_{\xi}^{\dagger} v_{N}\simeq ( \epsilon_{N}-h'_{\xi} ) ^{2}v_{N}.
\ee
There are terms in $h'_{\xi}$ and $h_{\xi}h_{\xi}^{\dagger}$ that do not commute with $\widehat{N}=\widehat a^{\dagger}\widehat a$. 
One solves (\ref{eq:secularequation}) by perturbation theory, that is justified by noting that the corrections are proportional to $a_{0}/\ell_{B}$ small parameter.
With this treatment, trigonal warping is taken into account at leading order. 
After some algebra, we can write the energies of the LLs in the large $\vert N\vert$ limit up to third-nearest neighbors hopping as
\be 
\epsilon_{N} = \epsilon_{N}^{(1)} + \epsilon_{N}^{(2)} + \epsilon_{N}^{(3)}.
\ee
Note that $\epsilon_{N}$ is independent of the valley $\xi$ index. The first two terms 
\begin{eqnarray}
  \epsilon_{N}^{(1)} & \approx & {\rm sgn}(N)\hslash\omega_{c} 
\vert N \vert^{1/2} \left(   1 - \dfrac{3}{8}  \dfrac{a_{0}^{2}}{\ell_B^{2}} \vert N \vert \right) ,
\\
\epsilon_{N}^{(2)} & \approx & \hslash\omega_{c} \dfrac{t^{(2)}}{\vert t^{(1)}\vert }\dfrac{3}{\sqrt{2}}\dfrac{a_{0}}{\ell_{B}} \vert N \vert \left(   1 - \dfrac{3}{4}  \dfrac{a_{0}^{2}}{\ell_B^{2}} \vert N \vert \right) ,
\end{eqnarray}
were obtained in Refs.~\cite{Plochocka2008,Goerbig2011}. Our derivation expands the latter results to account for $t^{(3)}$, namely
\be
\epsilon_{N}^{(3)} \approx -{\rm sgn}(N)\hslash\omega_{c} \dfrac{2t^{(3)}}{t^{(1)}} \vert N \vert^{1/2} \left( 1 - \dfrac{t^{(3)}}{t^{(1)}} - \dfrac{59}{32}  \dfrac{a_{0}^{2}}{\ell_B^{2}} \vert N \vert \right) .
\ee

\bibliography{hhk}
\end{document}